\documentclass[prl,aps,amsfonts,amssymb,amsmath,floats,floatfix,twocolumn,superscriptaddress,unsortedaddress]{revtex4}
\usepackage{graphics}
\usepackage{epsfig,graphicx}

\begin{document}


\title{Controlling the self-doping of YBa$_2$Cu$_3$O$_{7-\delta}$ polar surfaces:\\
From Fermi surface to nodal Fermi arcs by ARPES\\}

\author{M.A. Hossain}
\affiliation{Department of Physics {\rm {\&}} Astronomy, University of British Columbia, Vancouver, British Columbia V6T\,1Z1, Canada}
\author{J.D.F. Mottershead}
\affiliation{Department of Physics {\rm {\&}} Astronomy, University of British Columbia, Vancouver, British Columbia V6T\,1Z1, Canada}
\author{A. Bostwick}
\affiliation{Advanced Light Source, Lawrence Berkeley National Laboratory, Berkeley, California 94720, USA}
\author{J.L. McChesney}
\affiliation{Advanced Light Source, Lawrence Berkeley National Laboratory, Berkeley, California 94720, USA}
\author{E. Rotenberg}
\affiliation{Advanced Light Source, Lawrence Berkeley National Laboratory, Berkeley, California 94720, USA}
\author{\\R. Liang}
\affiliation{AMPEL, University of British Columbia, Vancouver, British Columbia V6T\,1Z4, Canada}
\author{W.N. Hardy}
\affiliation{Department of Physics {\rm {\&}} Astronomy, University of British Columbia, Vancouver, British
Columbia V6T\,1Z1, Canada} \affiliation{AMPEL, University of British Columbia, Vancouver, British Columbia
V6T\,1Z4, Canada}
\author{G.A. Sawatzky}
\affiliation{Department of Physics {\rm {\&}} Astronomy, University of British Columbia, Vancouver, British Columbia V6T\,1Z1, Canada}
\affiliation{AMPEL, University of British Columbia, Vancouver, British Columbia V6T\,1Z4, Canada}
\author{I.S. Elfimov}
\affiliation{AMPEL, University of British Columbia, Vancouver, British Columbia V6T\,1Z4, Canada}
\author{D.A. Bonn}
\affiliation{Department of Physics {\rm {\&}} Astronomy, University of British Columbia, Vancouver, British Columbia V6T\,1Z1, Canada}
\affiliation{AMPEL, University of British Columbia, Vancouver, British Columbia V6T\,1Z4, Canada}
\author{A. Damascelli}
\email{damascelli@physics.ubc.ca} \affiliation{Department of Physics {\rm {\&}} Astronomy, University of British Columbia, Vancouver, British
Columbia V6T\,1Z1, Canada} \affiliation{AMPEL, University of British Columbia, Vancouver, British Columbia V6T\,1Z4, Canada}

\maketitle

{\bf The discovery of quantum oscillations in the normal-state electrical resistivity of YBa$_2$Cu$_3$O$_{6.5}$ \cite{Louis2007} provides the
first evidence for the existence of Fermi surface (FS) pockets in an underdoped cuprate. However, the pockets' electron vs. hole character, and
the very interpretation in terms of closed FS contours, are the subject of considerable debate
\cite{yelland,nigel,greg,leboeuf,jaudet,ilya,carrington,harrisonPRL,rice,millis,sudip,alexandrov,vafek}. Angle-resolved photoemission
spectroscopy (ARPES), with its ability to probe electronic dispersion as well as the FS, is ideally suited to address this issue. Unfortunately,
the ARPES study of YBa$_2$Cu$_3$O$_{7-\delta}$ (YBCO) has been hampered by the technique's surface sensitivity \cite{schabel,lu,borisenko}. Here
we show that this stems from the polarity and corresponding self-doping of the YBCO surface. By in-situ deposition of potassium atoms on the
cleaved surface, we are able to continuously tune the doping of a single sample from the heavily overdoped to the underdoped regime. This
reveals the progressive collapse of the normal-metal-like FS into four disconnected nodal FS arcs, or perhaps into hole but not electron
pockets, in underdoped YBCO6.5.}

The key to the high-$T_c$ cuprate puzzle is understanding the evolution of the low-energy normal-state
electronic structure, upon doping charge carriers into the CuO$_2$ planes, from the underdoped
antiferromagnetic charge-transfer insulator to the overdoped normal metal. This sets the stage for the
emergence of high-$T_c$ superconductivity at intermediate doping. A spectacular account of this remarkable
evolution has been provided by the collapse - upon underdoping - of the normal-state FS. In the heavily
overdoped regime, angular magnetoresistance oscillation \cite{Hussey:2003} and ARPES experiments
\cite{29a,darren} on Tl$_2$Ba$_2$CuO$_{6+\delta}$ have arrived at a very precise quantitative agreement in
observing a coherent band-structure-like FS.
\begin{figure}[b]
\centerline{\epsfig{figure=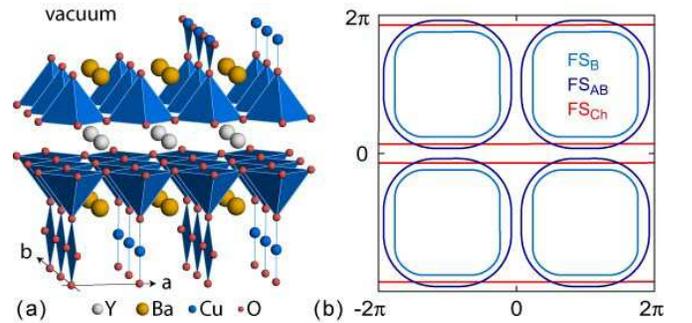,width=1\linewidth,clip=}} \caption{{\bf The surface of cleaved YBCO.} (a) Structure of
oxygen-ordered YBCO6.5 (with alternating oxygen full and empty chains), showing the BaO and CuO chain terminations of the cleaved surface.
Electronic reconstruction takes place at these polar surfaces, similar to the prototypical case of a polar catastrophe in ionic insulators with
a $|1+|1-|1+|1-|...$ layer-by-layer charge \cite{ronald,harold}. This leads to `overdoped-like' FS features for the surface topmost layers, as
shown pictorially in (b) for CuO$_2$-plane bonding and antibonding bands (FS$_{B}$ and FS$_{AB}$), and the one-dimensional CuO chain band
(FS$_{Ch}$).} \label{YBCO_PS}
\end{figure}
Upon reducing the number of holes in the CuO$_2$ planes, however, the hole FS volume decreases consistently with Luttinger's theorem until below
optimal doping, at which point the single-particle FS appears to reduce to four disconnected nodal Fermi arcs \cite{norman,kanigel,kmshen}. This
scenario was suggested from ARPES studies of Bi-cuprates \cite{norman,kanigel} and Ca$_{2-x}$Na$_x$CuO$_2$Cl$_2$ \cite{kmshen}, and is naturally
connected to the existence of the pseudogap.

The detection of quantum oscillations in oxygen-ordered ortho-II YBCO6.5 calls the Fermi arc scenario into
question, suggesting instead a FS reconstruction into either hole and/or electron pockets
\cite{Louis2007,leboeuf,jaudet}.
\begin{figure*}[t]
\centerline{\epsfig{figure=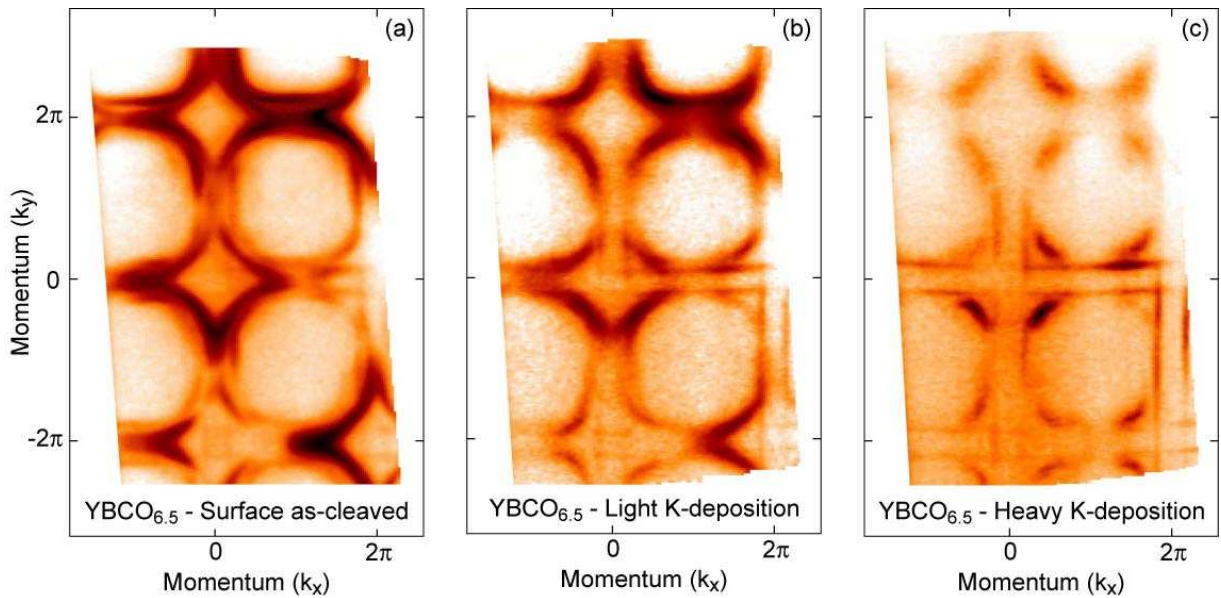,width=0.9\linewidth,clip=}} \caption{{\bf YBCO FS evolution upon $e^-$ doping.} (a) ARPES FS of
as-cleaved YBCO6.5, exhibiting an effective hole doping $p\!=\!0.28$ per planar Cu atom, as determined from the average area of bonding and
antibonding CuO$_2$-bilayer FSs. (b,c) By evaporating potassium on the same sample ($<\!1$ monolayer), electrons are transferred to the top-most
CuO$_2$ bilayer and the corresponding FSs become progressively more hole-underdoped. (c) For heavy K-deposition ($p\!=\!0.11$ as estimated from
the area of the chain FS), the $E_F$ ARPES intensity reduces to the 1D CuO chain FS and four disconnected nodal CuO$_2$ FS arcs.}
\label{YBCO_FS}
\end{figure*}
Due to the complex multiband and correlated character of the electronic structure of YBCO6.5 \cite{ilya}, the determination of the nature of
these pockets and their generality to the underdoped cuprates requires connecting transport and single-particle spectroscopy information on the
same underdoped system. The study of YBCO6.5 by ARPES is thus of extremely high priority. Unfortunately, this material is complicated by the
presence of CuO-chain layers and the lack of a neutral [001] cleavage plane (Fig.\,\ref{YBCO_PS}a). More specifically, YBCO cleaves between the
CuO chain layer and the BaO layer, leaving on the cleaved surface relatively large regions ($>$100\,\AA) of either CuO or BaO terminations.
Scanning-tunneling microscopy shows that the CuO chain terraces are characterized by prominent surface density waves \cite{8} and differ
substantially from the bulk: as also seen in ARPES \cite{schabel}, they exhibit surface states and unavoidable doping disorder. Recent ARPES
studies of nearly optimally doped YBCO indicated that CuO and BaO terminations give different contributions to the total photoemission intensity
\cite{borisenko}, with a doping $p\!=\!0.3$ for the top-most CuO$_2$ planes almost irrespective of the nominal bulk doping. This corresponds to
heavy overdoping all the way into the non-superconducting regime (Fig.\,\ref{YBCO_PD}), and is actually not achievable in bulk, fully-oxygenated
YBCO7.0 for which $p\!=\!0.194$ \cite{ruixing}. Similar problems have been encountered in the ARPES study of YBa$_2$Cu$_4$O$_8$ \cite{kaminski}.

Overcoming these problems requires, first of all, recognizing that the cleaved surface of YBCO is actually polar. This can lead to
overdoped-like FSs (Fig.\,\ref{YBCO_PS}b) even as a result of a pure electronic reconstruction \cite{ronald,harold}. A wide momentum
distribution map of the FS from `as-cleaved' YBCO6.5 is presented in Fig.\,\ref{YBCO_FS}a (Methods). The ARPES data are a superposition of
features from the BaO and CuO terminated regions: because of the few \AA\ probing depth at these photon energies, the ARPES intensity from the
BaO-terminated regions is dominated by the CuO$_2$-bilayer bands and that from the CuO-terminated regions by the chain band. The comparison with
Fig.\,\ref{YBCO_PS}b, allows one to identify the FS features originating from the CuO$_2$-plane bonding and antibonding bands (FS$_{B}$ and
FS$_{AB}$), and the one-dimensional (1D) CuO chain band (FS$_{Ch}$). Note that the strong momentum-dependent intensity modulation of the ARPES
features, which seems inconsistent with the sample symmetry, is simply a manifestation of the matrix elements effects associated with the
photon/crystal/electron geometry changing across the field of view (no symmetrization was performed). Also, this particular ortho-II sample
happened to be twinned, in the bulk and not just on the surface, as confirmed by x-ray diffraction (this has an effect on FS$_{Ch}$ but not on
the discussion of the four-fold symmetric Fermi arcs). Most importantly, the fit of the two-dimensional ARPES FSs over multiple zones returns
the following areas, counting electrons, relative to the Brillouin zone area $A_{BZ}\!=\!4\pi^2/ab$: the bonding Fermi surface area
FS$_B\!=\!46.2\%$, the antibonding FS$_{AB}\!=\!26.0\%$, and the chain surface FS$_{Ch}\!=\!13.8\%$. From the average of bonding and antibonding
FS areas, one can calculate the hole-doping $p\!=\!0.28\!\pm0.01$ per planar copper ($p\!=\!0$ for the 1/2-filled Mott insulator with 1 hole per
Cu atom). As summarized in the phase diagram of Fig.\,\ref{YBCO_PD}, this indicates that the self-doping of the YBCO6.5 polar surface is that of
a non-superconducting, heavily-overdoped cuprate.

The next step is that of actively controlling the self-doping of the surface, so as to reduce its hole content down to the value of underdoped,
bulk YBCO6.5. This can be achieved by evaporating potassium in situ on the cleaved surface (Methods): owing to the low ionization potential,
K$^{1+}$ ions are adsorbed on the surface and electrons are doped into the top-most layers. As a consequence, we would anticipate the evolution
of all detected features towards the underdoped regime of hole-doped cuprates, which is precisely what one can observe in Fig.\,\ref{YBCO_FS},
upon increasing the K$^{1+}$ concentration (i.e., decreasing the hole doping) from panel a to c. The doping is indeed changing according to an
increase in electron filling (all data were acquired on the same sample after subsequent K evaporations). This is evidenced by the continuous
FS$_{Ch}$ area increase (counting electrons), which evolves from FS$_{Ch}\!=\!13.8\%$ for the as-cleaved surface to FS$_{Ch}^{K1}\!=\!14.7\%$
and FS$_{Ch}^{K2}\!=\!16.6\%$ for the increasingly K-deposited YBCOK1 (Fig.\,\ref{YBCO_FS}b) and YBCOK2 (Fig.\,\ref{YBCO_FS}c).
\begin{figure}[t]
\centerline{\epsfig{figure=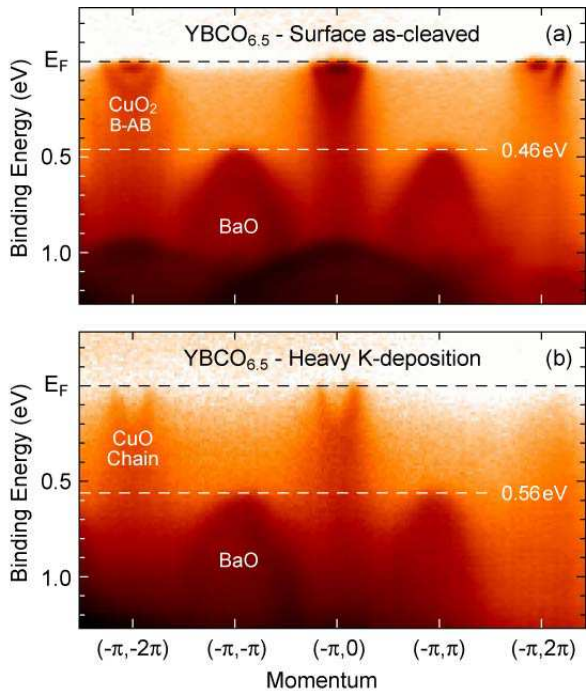,width=0.9\linewidth,clip=}} \caption{{\bf YBCO band evolution
upon $e^-$ doping.} (a) Electronic band dispersion as revealed by the $E$-vs.-$k$ ARPES intensity map for
as-cleaved YBCO6.5. (b) Upon K deposition, electrons are added as evidenced by the high binding energy shift
of the BaO bands; at the same time, the intensity of bonding (B) and antibonding (AB) CuO$_2$ features
vanishes and only the CuO chain band is detected at the antinodes.} \label{YBCO_BS}
\end{figure}
By comparing the carrier concentration per chain-Cu measured by the FS$_{Ch}$ area with the results of ab-initio LDA band structure calculations
\cite{ilya}, we can estimate the corresponding hole-doping per planar-Cu. This way we find good agreement with the value $p\!=\!0.28$ already
estimated for as-cleaved YBCO6.5 from the average of FS$_B$ and FS$_{AB}$. Most importantly, however, we obtain a hole-doping $p^{K1}\!=\!0.20$
and $p^{K2}\!=\!0.11$ for YBCOK1 and YBCOK2, which means that the surface of YBCOK2 is very close to the doping level $p\!=\!0.097$ of bulk
YBCO6.5.

The most interesting aspect of the data in Fig.\,\ref{YBCO_FS} is the evolution of the CuO$_2$-plane features. For heavy K deposition
(Fig.\,\ref{YBCO_FS}c), the LDA-like CuO$_2$-bilayer bonding and antibonding FSs of overdoped YBCO have collapsed into four nodal Fermi arcs,
consistent with other underdoped cuprates \cite{kmshen,norman,kanigel}. This is accompanied by the complete suppression of CuO$_2$ antinodal
spectral intensity as well as nodal bilayer splitting, which were instead clearly resolved for as-cleaved YBCO6.5 (Fig.\,\ref{YBCO_FS}a). Their
disappearance with K deposition hints at a severe loss of coherence upon underdoping. Correspondingly the CuO$_2$ nodal Fermi wavevectors,
relative to the Brillouin zone diagonal $(0,0)$-$(\pi,\pi)$, have evolved from $k_F^{AB}\!=\!0.29$ and $k_F^{B}\!=\!0.36$ for `overdoped'
as-cleaved YBCO6.5, to a single $k_F^{K2}\!=\!0.40$ for `underdoped' YBCOK2. These numbers compare well to what has been observed on other
cuprates at similar dopings (note that the following are both single CuO$_2$-layer systems): $k_F\!=\!0.36$ and 0.41, respectively, for
overdoped $p\!=\!0.26$ Tl$_2$Ba$_2$CuO$_{6+\delta}$ \cite{29a} and underdoped $p\!=\!0.12$ Ca$_{2-x}$Na$_x$CuO$_2$Cl$_2$ \cite{kmshen}.

The transfer of electrons to the surface of YBCO upon K deposition is also confirmed by an inspection of the electronic dispersions along the
$(-\pi,-2\pi)$-$(-\pi,2\pi)$ direction in momentum space. On as-cleaved YBCO6.5 (Fig.\,\ref{YBCO_BS}a), we detect well-defined bonding and
antibonding CuO$_2$ bands crossing $E_F$ at the antinodes, as well as the BaO band with a maximum binding energy at the zone corners
$(-\pi,\pm\pi)$. On YBCOK2 (Fig.\,\ref{YBCO_BS}b), the BaO band is now located $\sim\!100$\,meV deeper in energy. This indicates a shift of the
chemical potential consistent with the transfer of electrons from adsorbed K atoms to the top-most BaO, CuO-chain, and CuO$_2$-plane layers. At
the same time, on underdoped $p\!=\!0.11$ YBCOK2 the only coherent feature detected at the antinodes is the 1D CuO chain band; the antinodal
CuO$_2$-plane spectral weight has now become fully incoherent.

We would now like to summarize our findings, illustrated by the phase diagram and symmetrized FS data for YBCOK2 and as-cleaved YBCO6.5 in
Fig.\,\ref{YBCO_PD}. Our study demonstrates that the self-hole-doping of the cleaved polar surfaces of YBCO can be controlled by the in-situ
evaporation of alkali metals, in the present case K. This novel approach paves the way for the study of this important material family $-$
across the whole phase diagram $-$ by single particle spectroscopies. As this material has been the gold standard in a number of seminal
bulk-sensitive studies of the normal and superconducting properties, the direct connection with single-particle spectroscopy can lead to an
understandable underdoped anchor point, analogous to Tl$_2$Ba$_2$CuO$_{6+\delta}$ in the overdoped regime \cite{darren}. The results obtained
for $p\!=\!0.11$ YBCOK2 establish that the ARPES FS of underdoped YBCO consists of the superposition of 1D CuO chain FS and
CuO$_2$-plane-derived nodal Fermi arcs. It is thus consistent, with the additional complication of the chains, to what has already been observed
in oxychloride \cite{kmshen} and Bi-cuprates \cite {norman,kanigel}. In this sense, the disruption of the large hole-like coherent FS in
underdoped cuprates is a truly universal phenomenon.

Having obtained the first momentum resolved FS data for underdoped YBCO, it becomes crucial to understand the connection between ARPES and
quantum oscillation results \cite{Louis2007,jaudet}. First, the detection of the BaO band maximum at $\sim\!0.5$\,eV below $E_F$ does not
validate the scenario coming from LDA band structure calculations, which suggested that the small Fermi surface found in the quantum oscillation
measurements might originate from small pockets produced by BaO-Cu$_{chain}$ bands at $(\pm\pi,\pm\pi)$ \cite{ilya,carrington}. We also did not
observe any signature of CuO$_2$-derived band folding arising from the ortho-II oxygen-ordering of the chains \cite{ilya,carrington}, which is
possibly consistent with the loss of three-dimensional coherence evidenced by the suppression of bilayer band splitting upon underdoping.
\begin{figure}[t]
\centerline{\epsfig{figure=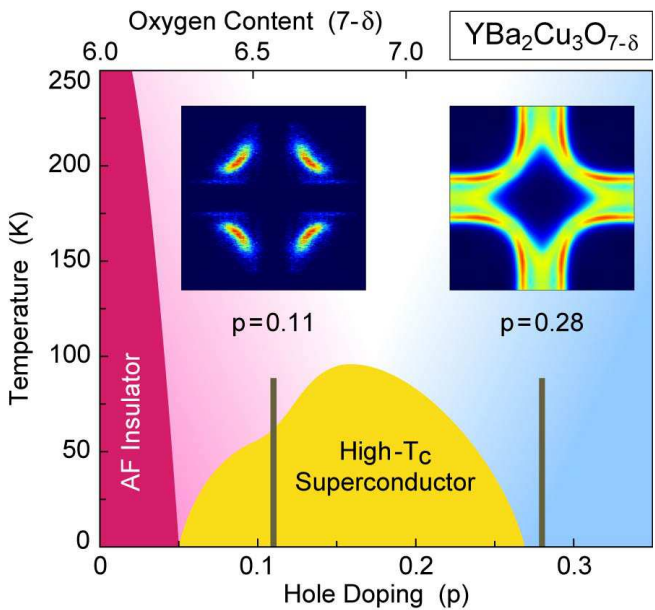,width=1.0\linewidth,clip=}} \caption{{\bf Phase diagram of YBCO by ARPES}. Schematic
temperature-doping phase diagram of YBCO adapted from Ref.\,\onlinecite{Louis2007}. The hole doping $p$ per planar copper ($p\!=\!0$ for the
1/2-filled Mott insulator with 1 hole per Cu atom), and the corresponding oxygen content ($7\!-\!\delta)$, are indicated on bottom and top
horizontal axes respectively \cite{ruixing}. The ARPES FS for under- and overdoped YBCO is also shown; the momentum-distribution maps have been
2-fold and 4-fold symmetrized for $p\!=\!0.11$ and 0.28, respectively. Similar to the data in Fig.\,\ref{YBCO_FS}, the doping levels were
determined for $p\!=\!0.11$ from the area of FS$_{Ch}$, and for $p\!=\!0.28$ from the area of FS$_{B}$ and FS$_{AB}$.} \label{YBCO_PD}
\end{figure}
Recent measurements of the Hall resistance in high magnetic field have noted a sign change with decreasing
temperature \cite{leboeuf}, suggesting that the quantum oscillations seen on top of a negative Hall
coefficient might come from small electron pockets, rather than the hole pockets originally proposed
\cite{Louis2007}. However, there is no sign of such electron pockets in our ARPES data from underdoped
YBCOK2, nor are there signs of additional zone-folding due to the kinds of density wave instabilities that
might give rise to such a Fermi surface reconstruction \cite{greg,leboeuf,millis}.

If any pocket had to be postulated on the basis of the present ARPES data, the most obvious possibility would be that the Fermi arcs are in fact
hole-like nodal Fermi pockets, as obtained for light doping of the antiferromagnetic parent compound in self-consistent Born calculations
\cite{sushkov} and already speculated from the study of other underdoped cuprates  \cite{kmshen,norman,kanigel}. The lack of a finite ARPES
intensity on the outer side of the pockets would be consistent with the strong drop in the quasiparticle coherence $Z_{\bf k}$ expected beyond
the antiferromagnetic zone boundary \cite{eder,harrisonPRL}. To estimate an area for these ostensible nodal pockets, we can fold the detected
arc profile, either with respect to the antiferromagnetic zone boundary or the end points of the arc itself, obtaining values of either 2.6\% or
1.3\%, relative to the full Brillouin zone area $A_{BZ}\!=\!4\pi^2/ab$. These numbers compare relatively well with the pocket area 1.9\%
suggested on the basis of quantum oscillation experiments on bulk YBCO6.5 \cite{Louis2007,jaudet}. However, these are hole, not electron
pockets.

At present, this seems to be a crucial disagreement between the interpretations of a single-particle spectroscopy and bulk transport
measurements on underdoped YBCO. The interpretation of the Hall resistance is likely to be complicated, since it necessarily invokes multiple
bands with different scattering and can also be influenced by strong correlations, the presence of localized magnetic moments, and also the
vortex liquid. It should also be noted that these transport measurements are made in high magnetic field whereas the ARPES data are acquired in
zero field. Nevertheless, there appears to be a discrepancy which must be resolved, especially since the YBCO ARPES data presented here are
consistent with the large body of single particle spectroscopy information obtained on other cuprates \cite{kmshen,norman,kanigel}.

Whatever the solution to the puzzle outlined above, it should be emphasized that the present approach, based
on the in-situ alkaline metal evaporation on freshly cleaved surfaces, opens the door to this type of
manipulation of other cuprates and complex oxides, not only to control the self-doping of polar surfaces but
also to reach doping levels otherwise precluded in the bulk. For instance, one could try to underdope the
surface of Tl$_2$Ba$_2$CuO$_{6+\delta}$, which grows naturally overdoped \cite{darren}, or even to obtain an
electron-doped superconductor starting from the insulating parent compounds.

\section{Methods}

\noindent {\bf Sample preparation.} YBa$_2$Cu$_3$O$_{6+x}$ single crystals were grown in non-reactive
BaZrO$_3$ crucibles using  a self-flux technique. The CuO$_x$ chain oxygen content was set to $x\!=\!0.51$ by
annealing in a flowing O$_2$:N$_2$ mixture and homogenized by further annealing in a sealed quartz ampoule,
together with ceramic at the same oxygen content. After mounting for the cleave required in an ARPES
measurement, the samples were cooled from 100\,$^{\circ}$C to room temperature over several days to establish
the ortho-II superstructure ordering of the CuO$_x$ chain layer \cite{ruixingPC}. The particular sample used
in the present study was twinned, as confirmed by x-ray diffraction after the ARPES measurements.\\

\noindent {\bf ARPES experiments.} ARPES measurements \cite{andreaPS} were carried out on the Electronic Structure Factory endstation at
Beamline 7.01 of the Advanced Light Source (ALS). The data were measured with linearly-polarized 110 eV photons and a Scienta R4000 electron
analyzer in angle-resolved mode. YBCO6.5 single crystals were cleaved in situ at a base pressure better than 2.5x10$^{-11}$\,torr and then
oriented by taking fast Fermi surface scans. Several procedures were tried out on these samples in an attempt to suppress the surface
contribution to the total photoemission intensity and gain direct insight into the bulk electronic structure. For instance, samples were
temperature-cycled between 20 and 100\,K \cite{andreaJESRP}, or were cleaved at higher temperature ($\sim\!80$\,K) and then cooled down to 20\,K
\cite{AndreaPRL}, in order to age and/or vary the characteristics of the cleaved surfaces. While both procedures have proved successful in
measuring the bulk dispersion and FS of layered Sr$_2$RuO$_4$ \cite{andreaJESRP,AndreaPRL}, no effect was observed in the case of YBCO6.5.
Successful control of the self-doping of the cleaved surfaces was achieved by in-situ deposition of submonolayers of potassium, with a
commercial SAES getter source \cite{aaron}, on freshly cleaved YBCO6.5. In this latter case, the samples were kept at 20\,K at all times during
the cleaving, K-deposition, and ARPES measurements. All ARPES data shown in Fig.\,\ref{YBCO_FS},\,\ref{YBCO_BS}, and\,\ref{YBCO_PD} were
obtained on the same sample, i.e. as-cleaved and after two subsequent K evaporations. Energy and angular resolutions were set to $\sim$\,30\,meV
and 0.2$^\circ$ ($\pm15^\circ$ angular window) for the data in Fig.\,\ref{YBCO_FS} and\,\ref{YBCO_BS}, and  to $\sim$\,30\,meV and 0.1$^\circ$
($\pm7^\circ$ angular window) for the higher-quality FS mappings in Fig.\,\ref{YBCO_PD}.

\bibliographystyle{plain}

\vspace{-0.2cm}
\section{Acknowledgments}
\vspace{-0.25cm}

This work was supported by the Alfred P. Sloan Foundation (A.D.), an ALS Doctoral Fellowship (M.A.H.), the
CRC Program (A.D. and G.A.S), NSERC, CFI, CIFAR Quantum Materials, and BCSI. The Advanced Light Source is
supported by the Director, Office of Science, Office of Basic Energy Sciences, of the U.S. Department of
Energy under Contract No. DE-AC02-05CH11231.

\vspace{-0.2cm}
\section{Authors Information}
\vspace{-0.25cm}

Correspondence and requests for materials should be addressed to A. Damascelli (damascelli@physics.ubc.ca).

\end{document}